\documentclass{amsart}
\usepackage[T1]{fontenc}
\newtheorem{theorem}{Theorem}
\usepackage{amssymb}
\usepackage{amsmath}
\usepackage{styles}
\usepackage{dblfloatfix}
\usepackage{hyperref}
\DeclareMathOperator\Det{det}
\usepackage[backend=biber]{biblatex}
\calclayout
\title{Submodularity of Mutual Information for Multivariate Gaussian Sources with Additive Noise}
\author{George Crowley, I\~naki Esnaola}
\date{\today}

\begin{document}

\begin{abstract}
Sensor placement approaches in networks often involve using information-theoretic measures such as entropy and mutual information. We prove that mutual information abides by submodularity and is non-decreasing when considering the mutual information between the states of the network and a subset of $k$ nodes subjected to additive white Gaussian noise. We prove this under the assumption that the states follow a non-degenerate multivariate Gaussian distribution.

\end{abstract} 

\maketitle

\section{Introduction}

A graph is characterized by the set of nodes $\Vc = \{1,2, \dots, n\}$ with $n \in \mathbb{N}$,  where each node corresponds to a system element, and the set of edges as $\Ec = \{(i,j) \in \Vc \times \Vc : \text{node} \ i  \ \text{is connected to node} \ j \}$, where each edge represents a connection between nodes in the network. Jointly, the set of edges $\Ec$ and the set of nodes $\Vc$ define an undirected graph $\Gc = \left(\Vc, \Ec \right)$. We assume that the state of the network can be described by the vector of random variables $X^n := (X_1,X_2,\dots,X_n)^{\transp}$. The observations obtained for a sensor placed at a node $i \in \Vc$ are denoted as $Y_i$ and are subject to i.i.d. additive white Gaussian noise (AWGN), denoted as formally as $Z_i \sim N(0,\sigma^2)$, with $\sigma \in \mathbb{R}_{+}$. Hence, the measurements obtained by the placed sensor $i$ is given by
    \begin{align} \label{AWGN}
        Y_i = X_i + Z_i, \quad i \in \Vc.
    \end{align}
    
Assuming that $k < n$ with $k \in \mathbb{N}$ sensors are placed in the network amongst $n$ nodes, then the observation vector $Y^k$ is defined as
\begin{align} \label{vec_AWGN}
    Y^{k} := (Y_{i_1}, \dots, Y_{i_k})^{\transp},
\end{align}

where the subscript $i_j$ denotes the $j$-th selected sensor. 
\begin{definition} \label{def_linear_obs}
The set of linear observation matrices 
is given by
\begin{equation}
 \label{H_k}
    \mathcal{H}_{k} := \bigcup_{\mathclap{\substack{\mathcal{A}\subseteq \Vc \\ |\mathcal{A}| = k}}}   \mathcal{H}_{k}(\mathcal{A}),
\end{equation}
with
\begin{equation}
 \label{H_k_aux}
 \mathcal{H}_{k}(\mathcal{A}) := \bigg\{\Hm \in \{0,1\}^{k \times n}: \Hm = \begin{pmatrix} 
\ev_{i_1}^{\transp},\ev_{i_2}^{\transp},\hdots,\ev_{i_k}^{\transp} \end{pmatrix}^{\transp}  \textnormal{where} \ i_j \in \mathcal{A} \subseteq \Vc \ \textnormal{for} \ j = 1,\dots,k \bigg\},
\end{equation}
where $\ev_i \in \{0,1\}^{n}$ is the $i$-th column basis vector, i.e. $1$ in the $i$-th position and $0$ otherwise. 
\end{definition}

Combining Definition \ref{def_linear_obs} with (\ref{vec_AWGN}) yields the following observation model:
\begin{align} \label{dam1}
    Y^{k} := \Hm X^n + Z^k, \quad  \text{for all} \ \Hm \in \mathcal{H}_{k}.
\end{align}

We consider the problem of finding the sensor placement $\mathcal{A} \subset \Vc$ such that we seek to extremize the optimization problem
\begin{align} \label{max_1}
  \Hm^*_k := \argmax_{\Hm \in \mathcal{H}_{k}}  \ I(X^n; \Hm X^n +  Z^{k}),
\end{align}

where $I(\cdot,\cdot)$ denotes the information-theoretic measure mutual information \cite{cover_elements_2005}. Further assuming that the probability distribution of the state variables satisfies $X^n \sim N_{n}(\muv, \Sigmam)$, where $\muv \in \mathbb{R}^n$ and $\Sigmam \in S_{++}^{n}$, then 
   \begin{align} 
        f(\Hm) := I(X^n; \Hm X^n +  Z^k) = \dfrac{1}{2} \log \left( \dfrac{1}{\sigma^{2k}} \Det\left( \Hm \Sigmam \Hm^{\transp} + \sigma^2 \Id_k \right) \right),  \quad \Hm \in \mathcal{H}_k,   \label{eq_max}
    \end{align}

where $\Det(\cdot)$ denotes the determinant of a square matrix, and $\Id_{k}$ denotes the $(k \times k)$ identity matrix. 

\begin{theorem}\label{th1}
Under the assumption $X^n \sim N_{n}(\muv, \Sigmam)$, where $\muv \in \mathbb{R}^n$ and $\Sigmam \in S_{++}^{n}$, the function $f\left(\Hm\right)$ satisfies the following properties:

\begin{enumerate}
    \item $f(\Hm)$ is 0 when $\Hm \in \mathcal{H}_{0}$.
    \item $f(\Hm)$ is submodular.
    \item $f(\Hm)$ is non-decreasing.
\end{enumerate}
\end{theorem}

Under the conditions of Theorem \ref{th1}, when the greedy heuristic is applied to the optimization problem posed in (\ref{max_1}), the heuristic always produces a solution whose value is at least $1 - \left(\frac{k-1}{k}\right)^{k}$ times the optimal value, which has a limiting value of $\left(\frac{e-1}{e}\right)$ \cite{nemhauser_analysis_1978}.

\section{Submodularity}

We begin by introducing the definitions of non-decreasing and submodular set functions.
\begin{definition}[Definition 2.1 \cite{nemhauser_analysis_1978}]
Given a finite set $\Omega$, a real-valued function $z$ on the set of subsets of $\Omega$ is called submodular if 
\begin{align}
    z(\mathcal{A}) + z(\mathcal{B}) \geq z(\mathcal{A} \cup \mathcal{B}) + z(\mathcal{A} \cap \mathcal{B}), \quad \forall \mathcal{A},\mathcal{B} \subseteq \Omega.
\end{align}

We shall often make use of the incremental value of adding element $j$ to the set $\mathcal{S}$, let $\rho_{j}(\mathcal{S}) = z(\mathcal{S} \cup \{ j\}) - z(\mathcal{S}).$
\end{definition}

\begin{proposition}[Proposition 2.1 \cite{nemhauser_analysis_1978}]
Each of the following statements is equivalent and defines a submodular set function. \label{prop:2.1}

\begin{itemize}
    \item[(i)] $z(\mathcal{A}) + z(\mathcal{B}) \geq z(\mathcal{A} \cup \mathcal{B}) + z(\mathcal{A} \cap \mathcal{B}), \quad \forall \mathcal{A},\mathcal{B} \subseteq \Omega.$
    \item[(ii)] $\rho_{j}(\mathcal{S}) \geq \rho_{j}(\mathcal{T}), \quad \forall \mathcal{S} \subseteq \mathcal{T} \subseteq \Omega, \quad \forall j \in \Omega \setminus \mathcal{T}.$
\end{itemize}
\end{proposition}

Condition \textit{(ii)} can be re-written as
\begin{align} \label{submodularity}
    z(\mathcal{S} \cup \{ j\}) - z(\mathcal{S}) \geq z(\mathcal{T} \cup \{ j\}) - z(\mathcal{T}), \quad \forall \mathcal{S} \subseteq \mathcal{T} \subseteq \Omega, \quad \forall j \in \Omega \setminus \mathcal{T}.
\end{align}

\begin{proposition}[Proposition 2.2 \cite{nemhauser_analysis_1978}] Each of the following statements is equivalent and defines a non-decreasing submodular set function.

\begin{itemize}
    \item[(i')] Submodularity: $z(\mathcal{A}) + z(\mathcal{B}) \geq z(\mathcal{A} \cup \mathcal{B}) + z(\mathcal{A} \cap \mathcal{B}), \quad \forall \mathcal{A},\mathcal{B} \subseteq \Omega.$
    \item[] Non-decreasing: $z(\mathcal{A}) \leq z(\mathcal{B}), \quad \forall \mathcal{A} \subseteq \mathcal{B} \subseteq \Omega.$
\end{itemize}
\end{proposition}

\section{Proof of Submodularity}

To keep the notation consistent, we translate the notation used in \cite{nemhauser_analysis_1978} to ours. Set $\Omega = \Vc$ and $\mathcal{S} := \{i_{\mathcal{S}_1},i_{\mathcal{S}_2},\dots,i_{\mathcal{S}_s}\}$ 
such that the cardinalty of $\mathcal{S} = s,$ with $\mathcal{S} \subseteq \Omega.$
Then, we can write our cost function $z(\mathcal{S})$ as
\begin{align}
    z(\mathcal{S}) = f\left(\Hm_{\mathcal{S}}\right) := \dfrac{1}{2} \log \left( \dfrac{1}{\sigma^{2s}} \Det\left( \Hm_{\mathcal{S}} \Sigmam \Hm_{\mathcal{S}}^{\transp} + \sigma^2 \Id_{s}\right) \right),
\end{align}

where the observation matrix $\Hm_{\mathcal{S}} = \begin{pmatrix} \ev_{i_{\mathcal{S}_1}}^{\transp},\ev_{i_{\mathcal{S}_2}}^{\transp},\hdots,\ev_{i_{\mathcal{S}_s}}^{\transp} \end{pmatrix}^{\transp}$. We will now prove conditions (1) - (3) from Theorem \ref{th1}.

\begin{proof}[Proof of condition (1)]
Let $\Hm \in \mathcal{H}_{0}$, then $I(X^n;Z^k) = 0$ since $Z^k$ are i.i.d. Gaussian random variables.
\end{proof}

Before proving condition (2), we first note some key results used throughout the proof. 

\begin{lemma}[Block matrix determinant property]\label{seber_block_det}

Denote the block matrix $\Mm$ as
\begin{align} \Mm :=
    \begin{pmatrix}
        \Am & \Bm \\ \Cm & \Dm
    \end{pmatrix}.
\end{align}

If $\Am$ is invertible \cite[Pg 290, 14.1]{seber_matrix_2007}, then (\ref{det_property_1}) holds. If $\Dm$ is invertible, then (\ref{det_property_2}) holds, where 
\begin{align} \label{det_property_1}
    \Det(\Mm) = \Det\begin{pmatrix}
        \Am & \Bm \\ \Cm & \Dm 
    \end{pmatrix} & = \Det(\Am) \Det(\Dm - \Cm \Am^{-1}\Bm) \\ & = \Det(\Dm)\Det(\Am - \Bm \Dm^{-1}\Cm).  \label{det_property_2}
\end{align}
\end{lemma}

\begin{lemma}[Block matrix inversion] \label{lemma_2}

Define $\Mm$ as in Lemma \ref{seber_block_det}. If the inverse of $\Mm$ exists, \cite[Pg 292-293,  14.10 (a, iv)]{seber_matrix_2007}, and $\Cm = \Bm^{\transp}$, then
    \begin{align}
    \Mm^{-1} =
    \begin{pmatrix}
        \Am & \Bm \\ \Bm^{\transp} & \Dm
    \end{pmatrix}^{-1} = \begin{pmatrix}
        \Am^{-1} & \zerov \\ \zerov & \zerov
    \end{pmatrix} + \begin{pmatrix}
        -\Am^{-1} \Bm \\ \Id_{\gamma}
    \end{pmatrix} \left(\Dm - \Bm^{\transp} \Am^{-1} \Bm \right)^{-1} \begin{pmatrix}
      -\Bm^{\transp} \Am^{-1},  & \Id_{\gamma}
    \end{pmatrix}.
\end{align}

\end{lemma}

\begin{lemma}\label{lemma_3}
Let $\Mm \succ 0$, and let $\Cm$ be $p \times n$ of rank $q$ ($q \leq p$) \cite[Pg 225, 10.31 (a)]{seber_matrix_2007}. Then:   
\begin{align}
    \Cm \Mm \Cm^{\transp} \succeq 0.
\end{align}
\end{lemma}

\begin{lemma}[Properties of symmetric positive definite matrices] \label{lemma_4}
Define the matrix $\Mm$ as in Lemma \ref{seber_block_det}. Further, assume that $\Mm$ is symmetric $(\Cm = \Bm^{\transp})$ \cite[14.26 (a)]{seber_matrix_2007}. Then the following statement holds: 

\begin{enumerate}
    \item[(a)] $\Mm \succ 0$ if and only if \ $(\iff)$ $\Am \succ 0$ and $\Dm - \Bm \Dm^{-1} \Bm^{\transp} \succ 0$.  
\end{enumerate}
    
\end{lemma}

\begin{lemma}[Determinant inequality] \label{lemma_5}
Suppose $\Am \succeq 0$ and $\Bm \succeq 0$ be $n \times n$ Hermitian matrices  \cite[10.59 (c)]{seber_matrix_2007}.  Then the following inequality holds:  \vspace{0mm}

\begin{itemize}
    \item[(c)] $\Det(\Am + \Bm) \geq \Det(\Am) + \Det(\Bm)$ with equality if and only if $\Am + \Bm$ is singular or $\Am = \zerov$ or $\Bm = \zerov$.
\end{itemize}
    
\end{lemma}

\begin{lemma}[Inverse of block matrices] \label{lemma_6}
Define the matrix $\Mm$ as in Lemma \ref{seber_block_det}. Suppose that $\Mm$ is non-singular and $\Dm$ is also non-singular \cite[14.11 (b)]{seber_matrix_2007}. Define $\Mm_{\Am \cdot \Dm} = \Am - \Bm \Dm^{-1} \Cm$, then
\begin{align}
    \Mm^{-1} = \begin{pmatrix}
     \Mm_{\Am \cdot \Dm}^{-1} & - \Mm_{\Am \cdot \Dm}^{-1} \Bm \Dm^{-1} \\ \\ - \Dm^{-1} \Cm \Mm_{\Am \cdot \Dm}^{-1} & \Dm^{-1} + \Dm^{-1} \Cm \Mm_{\Am \cdot \Dm}^{-1} \Bm \Dm^{-1}  
    \end{pmatrix}.
\end{align}
\end{lemma}

For the proof, we first note that $j \notin \mathcal{T}$, to match notation with (\ref{submodularity}), and $\mathcal{S} \subseteq \mathcal{T}$. We further make note of the following observation matrices:
 \begin{align}
       \Hm_{\{j\}} & = \begin{pmatrix} \ev_{j}^{\transp} \end{pmatrix}^{\transp}, \\
        \Hm_{\mathcal{S} \cup \{j\}} & = \begin{pmatrix} \ev_{i_{\mathcal{S}_1}}^{\transp},\ev_{i_{\mathcal{S}_2}}^{\transp},\hdots,\ev_{i_{\mathcal{S}_s}}^{\transp}, \ev_{{j}}^{\transp}
\end{pmatrix}^{\transp}. \end{align}

Assue there exists a set $\Gamma$ such that $\mathcal{S} \cup \Gamma = \mathcal{T}$. Note that if $\mathcal{S} = \mathcal{T}$, then the function is equal and hence submodular. Otherwise,   
 \begin{align} \Hm_{\Gamma} & = \begin{pmatrix} \ev_{i_{\Gamma_1}}^{\transp},\dots, \ev_{i_{\Gamma_{\gamma}}}^{\transp} \end{pmatrix}^{\transp}, \\
     \Hm_{\mathcal{T}} = \Hm_{\mathcal{S} \cup \Gamma} & = \begin{pmatrix} \ev_{i_{\mathcal{S}_1}}^{\transp},\ev_{i_{\mathcal{S}_2}}^{\transp},\hdots,\ev_{i_{\mathcal{S}_s}}^{\transp},\ev_{i_{\Gamma_1}}^{\transp},\dots, \ev_{i_{\Gamma_{\gamma}}}^{\transp} \end{pmatrix}^{\transp} \\ & \label{H_T} = \begin{pmatrix}
         \Hm_{\mathcal{S}} \\ \Hm_{\Gamma}
     \end{pmatrix}, \\
        \Hm_{\mathcal{T} \cup \{j\}} = \Hm_{\mathcal{S} \cup \Gamma \cup \{j\}} & = \begin{pmatrix} \ev_{i_{\mathcal{S}_1}}^{\transp},\ev_{i_{\mathcal{S}_2}}^{\transp},\hdots,\ev_{i_{\mathcal{S}_s}}^{\transp}, \ev_{i_{\Gamma_1}}^{\transp},\dots, \ev_{i_{\Gamma_{\gamma}}}^{\transp},\ev_{{j}}^{\transp} \end{pmatrix}^{\transp} \\ & = \begin{pmatrix}
            \Hm_{\mathcal{S}} \\ \Hm_{\Gamma} \\ \Hm_{\{j\}}
        \end{pmatrix}. \end{align}

The cardinality of each subset is denoted by: $|\Vc| = n,  \ |\Gamma| = \gamma$, \ $|\mathcal{T}| = s + \gamma = t$, and $|{\{j\}}| = 1.$ 

\begin{proof}[Proof of condition (2)]
From Proposition \ref{prop:2.1}, we need to show (with $\mathcal{S} \subseteq \mathcal{T}$, $j \notin \mathcal{T}$)
    \begin{align} \nonumber
          & \ \dfrac{1}{2} \log \left( \dfrac{1}{\sigma^{2 (s + 1)}} \Det\left( \Hm_{\mathcal{S} \cup \{j \}} \Sigmam \Hm_{\mathcal{S} \cup \{ j \}}^{\transp} + \sigma^2 \Id_{s+1}\right) \right) -   \dfrac{1}{2} \log \left( \dfrac{1}{\sigma^{2s}} \Det\left( \Hm_{\mathcal{S}} \Sigmam \Hm_{\mathcal{S}}^{\transp} + \sigma^2 \Id_{s} \right) \right) \\ \geq & \ \dfrac{1}{2} \log \left( \dfrac{1}{\sigma^{2 (t + 1)}} \Det\left( \Hm_{\mathcal{T} \cup \{ j \}} \Sigmam \Hm_{\mathcal{T} \cup \{ j \}}^{\transp} + \sigma^2 \Id_{t+1}\right) \right) -   \dfrac{1}{2} \log \left( \dfrac{1}{\sigma^{2t}} \Det\left( \Hm_{\mathcal{T}} \Sigmam \Hm_{\mathcal{T}}^{\transp} + \sigma^2 \Id_{t}\right) \right),  \nonumber
    \end{align}

    which can be simplified to
        \begin{align} \label{log_1}
        \log \left( \dfrac{\dfrac{1}{\sigma^{2}} \Det\left( \Hm_{\mathcal{S} \cup \{ j \}} \Sigmam \Hm_{\mathcal{S} \cup \{ j \}}^{\transp} + \sigma^2 \Id_{s+1}\right)}{\Det\left( \Hm_{\mathcal{S}} \Sigmam \Hm_{\mathcal{S}}^{\transp} + \sigma^2 \Id_{s} \right)} \right) \geq \log \left(\dfrac{\dfrac{1}{\sigma^{2}} \Det\left( \Hm_{\mathcal{T} \cup \{ j \}} \Sigmam \Hm_{\mathcal{T} \cup \{ j \}}^{\transp} + \sigma^2 \Id_{t+1}\right)}{\Det\left( \Hm_{\mathcal{T}} \Sigmam \Hm_{\mathcal{T}}^{\transp} + \sigma^2 \Id_{t}\right)} \right).
        \end{align}

         Since all determinant values are positive (confirmed by the assumption that $\Sigmam$ is positive definite) and $\log$ is a monotonic increasing function, (\ref{log_1}) becomes
         \begin{align} \nonumber
                \dfrac{\dfrac{1}{\sigma^{2}} \Det\left( \Hm_{\mathcal{S} \cup \{ j \}} \Sigmam \Hm_{\mathcal{S} \cup \{ j \}}^{\transp} + \sigma^2 \Id_{s+1}\right)}{\Det\left( \Hm_{\mathcal{S}} \Sigmam \Hm_{\mathcal{S}}^{\transp} + \sigma^2 \Id_{s}\right)} & \geq \dfrac{\dfrac{1}{\sigma^{2}} \Det\left( \Hm_{\mathcal{T} \cup \{ j \}} \Sigmam \Hm_{\mathcal{T} \cup \{ j \}}^{\transp} + \sigma^2 \Id_{t+1}\right)}{\Det\left( \Hm_{\mathcal{T}} \Sigmam \Hm_{\mathcal{T}}^{\transp} + \sigma^2 \Id_{t}\right)} \\ \implies
                 \dfrac{\Det\left( \Hm_{\mathcal{S} \cup \{ j \}} \Sigmam \Hm_{\mathcal{S} \cup \{ j \}}^{\transp} + \sigma^2 \Id_{s+1}\right)}{\Det\left( \Hm_{\mathcal{S}} \Sigmam \Hm_{\mathcal{S}}^{\transp} + \sigma^2 \Id_{s}\right)} & \geq \dfrac{\Det\left( \Hm_{\mathcal{T} \cup \{ j \}} \Sigmam \Hm_{T \cup \{ j\}}^{\transp} + \sigma^2 \Id_{t+1}\right)}{\Det\left( \Hm_{\mathcal{T}} \Sigmam \Hm_{\mathcal{T}}^{\transp} + \sigma^2 \Id_{t}\right)}. \label{det_calc_1} 
    \end{align}

Before proceeding, we notice that
  \begin{align}
         \Hm_{\mathcal{S} \cup \{ j \}} \Sigmam \Hm_{\mathcal{S} \cup \{ j \}}^{\transp} + \sigma^2 \Id_{s+1} = \begin{pmatrix} \Hm_{\mathcal{S}} \Sigmam \Hm_{\mathcal{S}}^{\transp} + \sigma^2 \Id_{s} & \Hm_{\mathcal{S}} \Sigmam \Hm_{\{j\}}^{\transp} \\ \\ \Hm_{\{j\}} \Sigmam \Hm_{\mathcal{S}}^{\transp} & \Hm_{\{j\}} \Sigmam \Hm_{\{j\}} + \sigma^2 \end{pmatrix},
    \end{align}

    and 
     \begin{align}
        \Hm_{\mathcal{S} \cup \Gamma \cup \{ j \}} \Sigmam \Hm_{\mathcal{S} \cup \Gamma \cup \{ j \}}^{\transp} + \sigma^2 \Id_{s + \gamma +1 } &  = \begin{pmatrix}
        \Hm_{\mathcal{T}} \Sigmam \Hm_{\mathcal{T}}^{\transp} + \sigma^2 \Id_{t} & \cov(\Hm_{\mathcal{T}} X^n, \Hm_{\{j\}} X^n) \\ \\ (\cov(\Hm_{\mathcal{T}} X^n, \Hm_{\{j\}} X^n)^{\transp} &  \Hm_{\{j\}} \Sigmam \Hm_{\{j\}}^{\transp} + \sigma^2    
        \end{pmatrix}.
    \end{align}

    The covariances can be calculated as
\begin{align}
  \cov\left(\Hm_{\mathcal{T}} X^n, \Hm_{\{j\}} X^n \right) & = \Hm_{\mathcal{T}} \cov\left(X^n,X^n\right)\Hm_{\{j\}}^{\transp} \\ & = \Hm_{\mathcal{T}} \Sigmam \Hm_{\{j\}}^{\transp},  \nonumber
\end{align}

and its transposition is
\begin{align}
    (\Hm_{\mathcal{T}} \Sigmam \Hm_{\{j\}}^{\transp})^{\transp} = \Hm_{\{j\}} \Sigmam \Hm_{\mathcal{T}}^{\transp}.
\end{align}

Then, using Lemma \ref{prop:2.1}, with $\Am = \Hm_{\mathcal{S}} \Sigmam \Hm_{\mathcal{S}}^{\transp} + \sigma^2 \Id_{s}$, $\Dm = \Hm_{\{j\}} \Sigmam \Hm_{\{j\}}^{\transp} + \sigma^2,
\Bm = \Hm_{\mathcal{S}} \Sigmam \Hm_{\{j\}}^{\transp}$, and $\Cm = \Hm_{\{j\}} \Sigmam \Hm_{\mathcal{S}}^{\transp})$, it follows that the left-hand side of (\ref{det_calc_1}) can be written as
\begin{align} \nonumber
    & = \dfrac{ \Det\left( \Hm_{\mathcal{S} \cup \{ j \}} \Sigmam \Hm_{\mathcal{S} \cup \{ j \}}^{\transp} + \sigma^2 \Id_{s+1} \right)}{\Det\left( \Hm_{\mathcal{S}} \Sigmam \Hm_{\mathcal{S}}^{\transp} + \sigma^2 \Id_{s}\right)} \\ \nonumber & = \dfrac{\Det\left( \Hm_{\mathcal{S}} \Sigmam \Hm_{\mathcal{S}}^{\transp} + \sigma^2 \Id_{s}\right) \det(\Hm_{\{j\}} \Sigmam \Hm_{\{j\}} + \sigma^2  - \Hm_{\{j\}} \Sigmam \Hm_{\mathcal{S}}^{\transp} \left(\Hm_{\mathcal{S}} \Sigmam \Hm_{\mathcal{S}}^{\transp} + \sigma^2 \Id_{s} \right)^{-1} \Hm_{\mathcal{S}} \Sigmam \Hm_{\{j\}}^{\transp}) }{\Det\left( \Hm_{\mathcal{S}} \Sigmam \Hm_{\mathcal{S}}^{\transp} + \sigma^2 \Id_{s}\right)} \\ & = \det\left(\Hm_{\{j\}} \Sigmam \Hm_{\{j\}}^{\transp} + \sigma^2  - \Hm_{\{j\}} \Sigmam \Hm_{\mathcal{S}}^{\transp} \left(\Hm_{\mathcal{S}} \Sigmam \Hm_{\mathcal{S}}^{\transp} + \sigma^2 \Id_{s} \right)^{-1} \Hm_{\mathcal{S}} \Sigmam \Hm_{\{j\}}^{\transp}\right). \label{1}
\end{align}

Using Lemma \ref{prop:2.1}, taking $\Am = \Hm_{\mathcal{T}} \Sigmam \Hm_{\mathcal{T}}^{\transp} + \sigma^2 \Id_{t}, \Dm = \Hm_{\{j\}} \Sigmam \Hm_{\{j\}}^{\transp} + \sigma^2, \Bm = \cov(\Hm_{\mathcal{T}} X^n, \Hm_{\{j\}} X^n)$, and $\Cm = \Bm^{\transp}$, it follows that the right-hand side of (\ref{det_calc_1}) can be written as
\begin{align} \nonumber
    & = \dfrac{\Det\left( \Hm_{\mathcal{T} \cup \{ j \}} \Sigmam \Hm_{\mathcal{T} \cup \{ j \}}^{\transp} + \sigma^2 \Id_{t+1}\right)}{\Det\left( \Hm_{\mathcal{T}} \Sigmam \Hm_{\mathcal{T}}^{\transp} + \sigma^2 \Id_{t}\right)} \\ \nonumber & = \dfrac{\Det\left( \Hm_{\mathcal{T}} \Sigmam \Hm_{\mathcal{T}}^{\transp} + \sigma^2 \Id_{t}\right) \Det\left( \Hm_{\{j\}} \Sigmam \Hm_{\{j\}}^{\transp} + \sigma^2 - \Hm_{\{j\}} \Sigmam \Hm_{\mathcal{T}}^{\transp} \left(\Hm_{\mathcal{T}} \Sigmam \Hm_{\mathcal{T}}^{\transp} + \sigma^2 \Id_{t} \right)^{-1} \Hm_{\mathcal{T}} \Sigmam \Hm_{\{j\}}^{\transp}\right)}{\Det\left( \Hm_{\mathcal{T}} \Sigmam \Hm_{\mathcal{T}}^{\transp} + \sigma^2 \Id_{t}\right)} \\ & = \Det\left( \Hm_{\{j\}} \Sigmam \Hm_{\{j\}}^{\transp} + \sigma^2 - \Hm_{\{j\}} \Sigmam \Hm_{\mathcal{T}}^{\transp} \left(\Hm_{\mathcal{T}} \Sigmam \Hm_{\mathcal{T}}^{\transp} + \sigma^2 \Id_{t} \right)^{-1} \Hm_{\mathcal{T}} \Sigmam \Hm_{\{j\}}^{\transp}\right). \label{2}
\end{align}

Since $\Sigmam \ \text{is} \ (n \times n), \ \Hm_{\{j\}} \ \text{is} \ (1 \times n), \ \Hm_{\mathcal{S}} \  \text{is} \ (s \times n), \ \Hm_{\mathcal{T}} \  \text{is} \ (t \times n)$, and hence $\Hm_{\{j\}} \Sigmam \Hm_{\mathcal{S}}^{\transp}$ is $(1 \times s)$, it follows that the resulting matrices inside the determinants of both (\ref{1}) and (\ref{2}) are scalars. Since the determinant of a scalar is just the scalar itself, this observation shows us that we can rewrite (\ref{det_calc_1}) as
\begin{align} \nonumber
   - &\Hm_{\{j\}} \Sigmam \Hm_{\mathcal{S}}^{\transp} \left(\Hm_{\{j\}} \Sigmam \Hm_{\mathcal{S}}^{\transp} + \sigma^2 \Id_{s} \right)^{-1} \Hm_{\mathcal{S}} \Sigmam \Hm_{\{j\}}^{\transp} \geq - \Hm_{\{j\}} \Sigmam \Hm_{\mathcal{T}}^{\transp} \left(\Hm_{\mathcal{T}} \Sigmam \Hm_{\mathcal{T}}^{\transp} + \sigma^2 \Id_{t} \right)^{-1} \Hm_{\mathcal{T}} \Sigmam \Hm_{\{j\}}^{\transp} \\ \nonumber
   \implies & \Hm_{\{j\}} \Sigmam \Hm_{\mathcal{S}}^{\transp} \left(\Hm_{\{j\}} \Sigmam \Hm_{\mathcal{S}}^{\transp} + \sigma^2 \Id_{s} \right)^{-1} \Hm_{\mathcal{S}} \Sigmam \Hm_{\{j\}}^{\transp}  \leq \Hm_{\{j\}} \Sigmam \Hm_{\mathcal{T}}^{\transp} \left(\Hm_{\mathcal{T}} \Sigmam \Hm_{\mathcal{T}}^{\transp} + \sigma^2 \Id_{t} \right)^{-1} \Hm_{\mathcal{T}} \Sigmam \Hm_{\{j\}}^{\transp} \\ \nonumber
   \implies & \Hm_{\{j\}} \Sigmam \Hm_{\mathcal{T}}^{\transp} \left(\Hm_{\mathcal{T}} \Sigmam \Hm_{\mathcal{T}}^{\transp} + \sigma^2 \Id_{t} \right)^{-1} \Hm_{\mathcal{T}} \Sigmam \Hm_{\{j\}}^{\transp} - \Hm_{\{j\}} \Sigmam \Hm_{\mathcal{S}}^{\transp} \left(\Hm_{\{j\}} \Sigmam \Hm_{\mathcal{S}}^{\transp} + \sigma^2 \Id_{s} \right)^{-1} \Hm_{\mathcal{S}} \Sigmam \Hm_{\{j\}}^{\transp} \geq 0 \\ 
   \implies & \Hm_{\{j\}} \Sigmam \Big( \Hm_{\mathcal{T}}^{\transp} \left(\Hm_{\mathcal{T}} \Sigmam \Hm_{\mathcal{T}}^{\transp} + \sigma^2 \Id_{t} \right)^{-1} \Hm_{\mathcal{T}} -  \Hm_{\mathcal{S}}^{\transp} \left(\Hm_{\{j\}} \Sigmam \Hm_{\mathcal{S}}^{\transp} + \sigma^2 \Id_{s} \right)^{-1} \Hm_{\mathcal{S}} \Big) \Sigmam \Hm_{\{j\}}^{\transp} \geq 0. \label{32}
\end{align} 

Using (\ref{H_T}) and (\ref{32}) yields
\begin{align}
    \Hm_{\{j\}} \Sigmam \Big(  \begin{pmatrix}
    \Hm_{\mathcal{S}}^{\transp}, & \Hm_{\Gamma}^{\transp}
\end{pmatrix} \left(\Hm_{\mathcal{T}} \Sigmam \Hm_{\mathcal{T}}^{\transp} + \sigma^2 \Id_{t} \right)^{-1}  \begin{pmatrix}
    \Hm_{\mathcal{S}} \\ \Hm_{\Gamma}
\end{pmatrix} & -  \Hm_{\mathcal{S}}^{\transp} \left(\Hm_{\mathcal{S}} \Sigmam \Hm_{\mathcal{S}}^{\transp} + \sigma^2 \Id_{s} \right)^{-1} \Hm_{\mathcal{S}} \Big) \Sigmam \Hm_{\{j\}}^{\transp} \geq 0. \label{f_35} \end{align}

Observe that we can further manipulate the inequality in (\ref{f_35}) to obtain
\begin{align}
    \nonumber \Hm_{\{j\}} \Sigmam \Big[  \begin{pmatrix}
    \Hm_{\mathcal{S}}^{\transp}, & \Hm_{\Gamma}^{\transp}
\end{pmatrix} \left(\Hm_{\mathcal{T}} \Sigmam \Hm_{\mathcal{T}}^{\transp} + \sigma^2 \Id_{t} \right)^{-1}  \begin{pmatrix}
    \Hm_{\mathcal{S}} \\ \Hm_{\Gamma}
\end{pmatrix} & -  \begin{pmatrix}\Hm_{\mathcal{S}}^{\transp}, & \Hm_{\Gamma}^{\transp}\end{pmatrix} \begin{pmatrix}\left(\Hm_{\mathcal{S}} \Sigmam \Hm_{\mathcal{S}}^{\transp} + \sigma^2 \Id_{s} \right)^{-1} & \zerov \\ \zerov & 0*\Id_{\gamma} \end{pmatrix} \begin{pmatrix}
   \Hm_{\mathcal{S}} \\ \Hm_{\Gamma}
\end{pmatrix} \Big] \Sigmam \Hm_{\{j\}}^{\transp} \geq 0. \end{align}

It then follows after using (\ref{H_T}) that 
\begin{align}\Hm_{\{j\}} \Sigmam   \Hm_{\mathcal{T}}^{\transp} \Bigg[ \left(\Hm_{\mathcal{T}} \Sigmam \Hm_{\mathcal{T}}^{\transp} + \sigma^2 \Id_{t} \right)^{-1} & -  \begin{pmatrix}\left(\Hm_{\mathcal{S}} \Sigmam \Hm_{\mathcal{S}}^{\transp} + \sigma^2 \Id_{s} \right)^{-1} & \zerov \\ \zerov^{\transp} & 0*\Id_{\gamma} \end{pmatrix} \Bigg] \Hm_{\mathcal{T}} \Sigmam \Hm_{\{j\}}^{\transp} \geq 0.
\end{align}

The inequality holds if the matrix inside is positive semi-definite, i.e.
\begin{align} \label{38}
    \Bigg( \left( \Hm_{\mathcal{T}} \Sigmam \Hm_{\mathcal{T}}^{\transp} + \sigma^2 \Id_{t} \right)^{-1} & -  \begin{pmatrix}\left( \Hm_{\mathcal{S}} \Sigmam \Hm_{\mathcal{S}}^{\transp} + \sigma^2 \Id_{s} \right)^{-1} & \zerov \\ \zerov^{\transp} & 0*\Id_{\gamma} \end{pmatrix} \Bigg) \succeq 0.
\end{align}

The block form of $\Hm_{\mathcal{T}} \Sigmam \Hm_{\mathcal{T}}^{\transp} + \sigma^2 \Id_{t}$ can be expressed as
\begin{align} \label{39}
     \Hm_{\mathcal{S} \cup \Gamma} \Sigmam \Hm_{\mathcal{S} \cup \Gamma}^{\transp} + \sigma^2 \Id_{s + \gamma} & = \begin{pmatrix}
        \Hm_{\mathcal{S}} \Sigmam \Hm_{\mathcal{S}}^{\transp} + \sigma^2 \Id_{s} & \cov(\Hm_{\mathcal{S}} X^n, \Hm_{\Gamma} X^n) &  \\ \\ \left( \cov(\Hm_{\mathcal{S}} X^n, \Hm_{\Gamma} X^n) \right)^{\transp} &  \Hm_{\Gamma} \Sigmam \Hm_{\Gamma}^{\transp} + \sigma^2 \Id_{\gamma} 
        \end{pmatrix} = \begin{pmatrix}
            \Am & \Bm \\ \\ \Bm^{\transp} & \Cm
        \end{pmatrix}.
\end{align}

Using Lemma \ref{2}, with $\Am = \Hm_{\mathcal{S}} \Sigmam \Hm_{\mathcal{S}}^{\transp} + \sigma^2 \Id_{s}$, $\Bm$ and $\Dm$ as indicated from (\ref{39}), it follows that 
\begin{align}
    \left( \Hm_{S \cup \Gamma } \Sigmam \Hm_{\mathcal{S} \cup \Gamma}^{\transp} + \sigma^2 \Id_{s + \gamma} \right)^{-1} = \begin{pmatrix}
        \Am^{-1} & \zerov \\ \zerov & \zerov
    \end{pmatrix} + \begin{pmatrix}
        -\Am^{-1} \Bm \\ \Id_{\gamma}
    \end{pmatrix} \left(\Dm - \Bm^{\transp} \Am^{-1} \Bm \right)^{-1} \begin{pmatrix}
      -\Bm^{\transp} \Am^{-1},  & \Id_{\gamma}
    \end{pmatrix}. \label{34}
\end{align}

Inserting equation (\ref{34}) into (\ref{38}) yields the condition
\begin{align} \label{36}
\begin{pmatrix}
        -\Am^{-1} \Bm \\ \Id_{\gamma}
    \end{pmatrix} \left(\Dm - \Bm^{\transp} \Am^{-1} \Bm \right)^{-1} \begin{pmatrix}
      -\Bm^{\transp} \Am^{-1},  & \Id_{\gamma}
    \end{pmatrix} \succeq 0.
\end{align}

Observe that $\Am = \Hm_{\mathcal{S}} \Sigmam \Hm_{\mathcal{S}}^{\transp} + \sigma^2 \Id_{s}$ is symmetric and positive definite, then it follows that $\Am^{-1}$ is also symmetric and positive definite (i.e. $\Am \succ 0$, and $\left(\Am^{-1}\right)^{\transp} = \Am^{-1}$). Then it follows that
\begin{align}
    \begin{pmatrix}
        -\Am^{-1} \Bm \\ \Id_{\gamma}
    \end{pmatrix}^{\transp} = \begin{pmatrix}
        \left(-\Am^{-1} \Bm\right)^{\transp}, & \Id_{\gamma} \end{pmatrix} = \begin{pmatrix}
        -\Bm^{\transp} \Am^{-1}, & \Id_{\gamma}
    \end{pmatrix}.
\end{align}

By setting
\begin{align}
    \Cm := \begin{pmatrix}
        -\Am^{-1} \Bm \\ \Id_{\gamma}
    \end{pmatrix},
\end{align}

and using Lemma 3, it follows that the inequality in (\ref{36}) can be written as
\begin{align} \label{sub_final}
   \Cm \left(\Dm - \Bm^{\transp} \Am^{-1} \Bm \right)^{-1} \Cm^{\transp} \succeq 0 \iff \left(\Dm - \Bm^{\transp} \Am^{-1} \Bm \right)^{-1} \succ 0 \iff \Dm - \Bm^{\transp} \Am^{-1} \Bm \succ 0.
\end{align}

Moreover, by setting $\Wm := \Hm_{\mathcal{S} \cup \Gamma } \Sigmam \Hm_{\mathcal{S} \cup \Gamma}^{\transp} + \sigma^2 \Id_{s + \gamma}$ as in (\ref{39}), which is positive definite, by Lemma \ref{lemma_4}, it follows that $\Wm$ is positive definite if and only if $\Am \succ 0$ and $\Dm - \Bm^{\transp} \Am^{-1} \Bm \succ 0$. But $\Dm - \Bm^{\transp} \Am^{-1} \Bm \succ 0$ is the inequality in (\ref{sub_final}), and so the result follows.

\end{proof}

\begin{proof}[Proof of condition (3)]

Using the same notation as before, the non-decreasing property states
\begin{align} \label{non-decreasing}
    z(\mathcal{S}) \leq z(\mathcal{T}), \quad \forall \mathcal{S} \subseteq \mathcal{T} \subseteq \Vc.
\end{align}

In our formulation, the non-decreasing property yields as 
\begin{align}
    \dfrac{1}{2} \log \left( \dfrac{1}{\sigma^{2s}} \Det\left( \Hm_{\mathcal{S}} \Sigmam \Hm_{\mathcal{S}}^{\transp} + \sigma^2 \Id_{s} \right) \right) \leq \dfrac{1}{2} \log \left( \dfrac{1}{\sigma^{2t}} \Det\left( \Hm_{\mathcal{T}} \Sigmam \Hm_{\mathcal{T}}^{\transp} + \sigma^2 \Id_{t} \right) \right).
\end{align}

First, let us assume that $\mathcal{S} = \mathcal{T}$, then the equality holds trivially. Hence, we assume that $\mathcal{T} =\mathcal{S}\cup \Gamma$, then using the monotonicity of the logarithm, it follows that
\begin{align} \label{decreasing_ineq1}
      \dfrac{1}{\sigma^{2s}} \Det\left( \Hm_{\mathcal{S}} \Sigmam \Hm_{\mathcal{S}}^{\transp} + \sigma^2 \Id_{s} \right) \leq \dfrac{1}{\sigma^{2t}} \Det\left( \Hm_{\mathcal{T}} \Sigmam \Hm_{\mathcal{T}}^{\transp} + \sigma^2 \Id_{t} \right).
\end{align}

We set the block matrix $\Mm$ as
\begin{align}
       \Mm = \Hm_{\mathcal{T}} \Sigmam \Hm_{\mathcal{T}}^{\transp} + \sigma^2 \Id_{t} & = \begin{pmatrix}
        \Hm_{\mathcal{S}} \Sigmam \Hm_{\mathcal{S}}^{\transp} + \sigma^2 \Id_{s} & \cov(\Hm_{\mathcal{S}} X^n, \Hm_{\Gamma} X^n) \\ \\ \left( \cov(\Hm_{\mathcal{S}} X^n, \Hm_{\Gamma} X^n) \right)^{\transp} &  \Hm_{\Gamma} \Sigmam \Hm_{\Gamma}^{\transp} + \sigma^2 \Id_{\gamma}
        \end{pmatrix} = \begin{pmatrix}
            \Am & \Bm \\ \Cm & \Dm
        \end{pmatrix},
\end{align}

then, by Lemma \ref{seber_block_det}, it follows that
\begin{align}
  \det(\Mm) & = \det(\Am) \det(\Dm - \Cm \Am^{-1}\Bm) \\ & = \det\left(\Hm_{\mathcal{S}} \Sigmam \Hm_{\mathcal{S}}^{\transp} + \sigma^2 \Id_{s}\right) \det(\Dm - \Cm \Am^{-1}\Bm). \label{det_ineq_1}
\end{align}

Using (\ref{det_ineq_1}) in (\ref{decreasing_ineq1}) yields
\begin{align} \label{decreasing_ineq2}
    \dfrac{1}{\sigma^{2s}} \det\left( \Hm_{\mathcal{S}} \Sigmam \Hm_{\mathcal{S}}^{\transp} + \sigma^2 \Id_{s} \right) \leq \dfrac{1}{\sigma^{2t}} \det\left(\Hm_{\mathcal{S}} \Sigmam \Hm_{\mathcal{S}}^{\transp} + \sigma^2 \Id_{s}\right) \det(\Dm - \Cm \Am^{-1}\Bm).
\end{align}

Since $\Hm_{\mathcal{S}} \Sigmam \Hm_{\mathcal{S}}^{\transp} + \sigma^2 \Id_{s} \succ 0 \implies \det\left( \Hm_{\mathcal{S}} \Sigmam \Hm_{\mathcal{S}}^{\transp} + \sigma^2 \Id_{s} \right) > 0$, we can divide this term out of (\ref{decreasing_ineq2}) such that
\begin{align} \label{decreasing_ineq3}
    \dfrac{1}{\sigma^{2s}} \leq \dfrac{1}{\sigma^{2t}} \det(\Dm - \Cm \Am^{-1}\Bm),
\end{align}

and hence, using $t = s + \gamma$ and fully expanding all the terms, (\ref{decreasing_ineq3}) can be written as 
\begin{align} \label{48}
   \det \left( \Hm_{\Gamma} \Sigmam \Hm_{\Gamma}^{\transp} + \sigma^2 \Id_{\gamma} - \left( \cov(\Hm_{\mathcal{S}} X^n, \Hm_{\Gamma} X^n) \right)^{\transp} \left( \Hm_{\mathcal{S}} \Sigmam \Hm_{\mathcal{S}}^{\transp} + \sigma^2 \Id_{s}\right)^{-1}  \cov(\Hm_{\mathcal{S}} X^n, \Hm_{\Gamma} X^n) \right) \geq \sigma^{2\gamma}.
\end{align}

Set $\Am = \sigma^2 \Id_{\gamma}$ and $\Bm = \Hm_{\Gamma} \Sigmam \Hm_{\Gamma}^{\transp} - \left( \cov(\Hm_{\mathcal{S}} X^n, \Hm_{\Gamma} X^n) \right)^{\transp} \left( \Hm_{\mathcal{S}} \Sigmam \Hm_{\mathcal{S}}^{\transp} + \sigma^2 \Id_{s}\right)^{-1}  \cov(\Hm_{\mathcal{S}} X^n, \Hm_{\Gamma} X^n)$. We omit temporarily showing that $\Bm \succeq 0$, but will invoke Lemma \ref{lemma_5} on (\ref{48}) which yields the inequality
\begin{align}
    \det(\Am+\Bm) \geq \det(\Am) + \det(\Bm) \geq \sigma^{2\gamma}.\end{align}

Since $\Am = \sigma^2 \Id_{\gamma}$, we have $\det(\Am) = \sigma^{2\gamma}$. Then
\begin{align}
    \det(\Am+\Bm) \geq \sigma^{2\gamma} + \det(\Bm) \geq \sigma^{2\gamma} \implies \det(\Bm) \geq 0 \iff \Bm \succeq 0.
\end{align}

We will now proceed by showing that $\Bm$ is semi-positive definite. We can write the joint random vector of $\Hm_{\Gamma} X^n$ and $\Hm_{\mathcal{S}} X^n + Z^{s}$ as
\begin{align}
    \begin{pmatrix}
        \Hm_{\Gamma} X^n \\ \Hm_{\mathcal{S}} X^n + Z^{s}
    \end{pmatrix} & \sim N\left( \begin{pmatrix}
        \Hm_{\Gamma} \E[X^n] \\ \Hm_{\mathcal{S}} \E[X^n]
    \end{pmatrix}, \begin{pmatrix}
        \cov\left(\Hm_{\Gamma} X^n,\Hm_{\Gamma} X^n\right) & \cov\left(\Hm_{\Gamma} X^n,\Hm_{\mathcal{S}} X^n + Z^{s}\right) \\ \cov\left(\Hm_{\mathcal{S}} X^n + Z^{s},\Hm_{\Gamma} X^n\right) & \cov\left(\Hm_{\mathcal{S}} X^n + Z^{s},\Hm_{\mathcal{S}} X^n + Z^{s}\right)
    \end{pmatrix} \right) \\ & \sim N\left( \begin{pmatrix}
        \Hm_{\Gamma} \E[X^n] \\ \Hm_{\mathcal{S}} \E[X^n]
    \end{pmatrix}, \begin{pmatrix}
       \Hm_{\Gamma}  \Sigmam \Hm_{\Gamma}^{\transp} & \Hm_{\Gamma} \Sigmam \Hm_{\mathcal{S}}^{\transp} \\ \Hm_{\mathcal{S}} \Sigmam \Hm_{\Gamma}^{\transp} & \Hm_{\mathcal{S}}  \Sigmam \Hm_{\mathcal{S}}^{\transp} + \sigma^2 \Id_{s}
    \end{pmatrix} \right). \label{53}
\end{align}

Observe that the covariance matrix in (\ref{53}) is positive definite, since
\begin{align}
    \begin{pmatrix}\label{55}
       \Hm_{\Gamma}  \Sigmam \Hm_{\Gamma}^{\transp} & \Hm_{\Gamma} \Sigmam \Hm_{\mathcal{S}}^{\transp} \\ \Hm_{\mathcal{S}} \Sigmam \Hm_{\Gamma}^{\transp} & \Hm_{\mathcal{S}}  \Sigmam \Hm_{\mathcal{S}}^{\transp} + \sigma^2 \Id_{s}
    \end{pmatrix} = \begin{pmatrix}
       \Hm_{\Gamma}  \Sigmam \Hm_{\Gamma}^{\transp} & \Hm_{\Gamma} \Sigmam \Hm_{\mathcal{S}}^{\transp} \\ \Hm_{\mathcal{S}} \Sigmam \Hm_{\Gamma}^{\transp} & \Hm_{\mathcal{S}}  \Sigmam \Hm_{\mathcal{S}}^{\transp}
    \end{pmatrix} + \begin{pmatrix}
        \zerov_{\gamma \times \gamma} & \zerov \\ \zerov^{\transp} & \sigma^2 \Id_s
    \end{pmatrix},
\end{align}

and the first matrix is a principle submatrix of $\Sigmam$, which is positive definite by assumption. Hence, the inverse of the covariance matrix in (\ref{55}) exists, which is also positive definite. By Lemma \ref{lemma_6}, it then follows that
\begin{align}
      \begin{pmatrix}
       \Hm_{\Gamma}  \Sigmam \Hm_{\Gamma}^{\transp} & \Hm_{\Gamma} \Sigmam \Hm_{\mathcal{S}}^{\transp} \\ \Hm_{\mathcal{S}} \Sigmam \Hm_{\Gamma}^{\transp} & \Hm_{\mathcal{S}}  \Sigmam \Hm_{\mathcal{S}}^{\transp} + \sigma^2 \Id_{s}
    \end{pmatrix}^{-1} = \begin{pmatrix}
    \left(\Hm_{\Gamma} \Sigmam \Hm_{\Gamma}^{\transp} - \Hm_{\Gamma} \Sigmam \Hm_{\mathcal{S}}^{\transp}\left(\Hm_{\mathcal{S}}  \Sigmam \Hm_{\mathcal{S}}^{\transp} + \sigma^2 \Id_{s}\right)^{-1}\Hm_{\mathcal{S}} \Sigmam \Hm_{\Gamma}^{\transp} \right)^{-1}  & \dots \\ \\
     \dots   & \dots
    \end{pmatrix}.
\end{align}

Since the covariance matrix is positive definite, Lemma \ref{lemma_4} implies that 
\begin{align} \left(\Hm_{\Gamma} \Sigmam \Hm_{\Gamma}^{\transp} - \Hm_{\Gamma} \Sigmam \Hm_{\mathcal{S}}^{\transp}\left(\Hm_{\mathcal{S}}  \Sigmam \Hm_{\mathcal{S}}^{\transp} + \sigma^2 \Id_{s}\right)^{-1}\Hm_{\mathcal{S}} \Sigmam \Hm_{\Gamma}^{\transp} \right)^{-1} & \succ 0 \\ \iff \Hm_{\Gamma} \Sigmam \Hm_{\Gamma}^{\transp} - \Hm_{\Gamma} \Sigmam \Hm_{\mathcal{S}}^{\transp}\left(\Hm_{\mathcal{S}}  \Sigmam \Hm_{\mathcal{S}}^{\transp} + \sigma^2 \Id_{s}\right)^{-1}\Hm_{\mathcal{S}} \Sigmam \Hm_{\Gamma}^{\transp} & \succ 0. \label{57} \end{align}

But the matrix in (\ref{57}) is $\Bm$, since $\left( \cov(\Hm_{\mathcal{S}} X^n, \Hm_{\Gamma} X^n) \right)^{\transp} = \left(\Hm_{\mathcal{S}} \Sigmam \Hm_{\Gamma}^{\transp} \right)^{\transp} = \Hm_{\Gamma} \Sigmam \Hm_{\mathcal{S}}^{\transp}$, and hence the result follows.   
\end{proof}

\printbibliography[]
\end{document}